\documentclass[aps,prl,twocolumn,superscriptaddress,showpacs]{revtex4}

\usepackage{graphicx,color,amsmath}

\begin{document}

\title{
First-Principles Study for the Anisotropy of Iron-based Superconductors toward Power and Device Applications
}

\author{Hiroki Nakamura}
\affiliation{CCSE, Japan Atomic Energy Agency, 6--9--3 Higashi-Ueno,
Taito-ku Tokyo 110--0015, Japan}
\affiliation{JST, Transformative Research-Project on Iron Pnictides (TRIP), Chiyoda, Tokyo 102-0075, Japan}
\affiliation{CREST (JST), 4--1--8 Honcho, Kawaguchi, Saitama 332--0012,
Japan}
\author{Masahiko Machida}
\affiliation{CCSE, Japan Atomic Energy Agency, 6--9--3 Higashi-Ueno,
Taito-ku Tokyo 110--0015, Japan}
\affiliation{JST, Transformative Research-Project on Iron Pnictides (TRIP), Chiyoda, Tokyo 102-0075, Japan}
\affiliation{CREST (JST), 4--1--8 Honcho, Kawaguchi, Saitama 332--0012,
Japan}
\author{Tomio Koyama}
\affiliation{Institute for Materials Research, Tohoku University, 2-1-1 Katahira, Aoba-ku, Sendai 980-8577, Japan}
\affiliation{CREST (JST), 4--1--8 Honcho, Kawaguchi, Saitama 332--0012,
Japan}
\author{Noriaki Hamada}
\affiliation{Faculty of Science and Technology, Tokyo University of Science, 2641 Yamazaki, Noda 278-8510, Japan  }

\date{\today}

\begin{abstract}

Performing the first-principles calculations, we investigate the anisotropy in the superconducting 
state of iron-based superconductors to gain an insight into their potential applications. The 
anisotropy ratio $\gamma_\lambda$ of the $c$-axis penetration depth to the $ab$-plane one is relatively 
small in BaFe$_2$As$_2$ and LiFeAs, i.e., $\gamma_\lambda \sim 3$, indicating that the transport 
applications are promising in these superconductors. On the other hand, in those having perovskite 
type blocking layers such as Sr$_2$ScFePO$_3$ we find a very large value, 
$\gamma_\lambda \gtrsim 200$, comparable to that in strongly anisotropic high-$T_c$ cuprate 
Bi$_2$Sr$_2$CaCu$_2$O$_{8-\delta}$. Thus, the intrinsic Josephson junction stacks are expected to 
be formed along the $c$-axis, and novel Josephson effects due to the multi-gap nature are also suggested 
in these superconductors.

\end{abstract}


\maketitle

\begin{figure}
\includegraphics[scale=0.13]{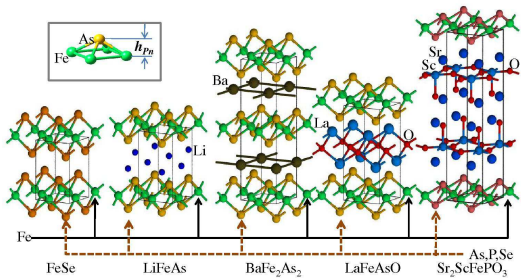}
\caption{Crystal structures of typical iron-based superconductors.}\label{fig:crystal}
\end{figure}

Since the discovery of iron-based  superconductor LaFeAsO$_{1-x}$F$_x$ with $T_{\rm c}=26$K by 
Kamihara {\it et al.}\cite{kamihara}, high-$T_{\rm c}$ superconductivity has been reported  
in some of its family materials. All the superconductors in this family have FePn layers (Pn = P, As, or Se), 
in which the high-$T_{\rm c}$ superconductivity emerges, and  are classified  into five groups 
according to the non-superconducting blocking layers sandwiched between FePn layers 
as seen in Fig.~\ref{fig:crystal}, 
i.e., 1111 (e.g., LaFeAsO), 122 (e.g., BaFe$_2$As$_2$)\cite{122}, 111(e.g., LiFeAs)\cite{111}, 
11(e.g., FeSe)\cite{11}, and the other type that contains very thick perovskite-based blocking 
layers (Sr$_2$ScFePO$_3$)\cite{ogino}.

The quasi two-dimensionality of the electronic states due to the layer structure of the iron-based 
superconductors is significant for their applications, since it causes anisotropy in superconducting 
properties such as the critical current $J_c$ and the upper and lower critical fields, $H_{c2}$ and $H_{c1}$.
A convenient quantity characterizing the 
superconducting anisotropy is the anisotropy parameter $\gamma_\lambda$ defined as the ratio of 
the $c$-axis penetration depth to the in-plane one. 
 From the values of $\gamma_\lambda$, one can extract important information about their potential 
applications. For example, superconductors with small $\gamma_\lambda$ are suitable for power 
applications using superconducting wires, tapes or cables. On the other hand, those with extremely 
large $\gamma_\lambda$ are expected to work as the intrinsic Josephson junctions \cite{kleiner,oya}, 
which will be promising as a THz emission device \cite{ozyuzer,kadowaki}.

It is well known that the penetration depth anisotropy $\gamma_\lambda$ directly leads to the anisotropy 
of $H_{c2}$, $H_{c1}$ and $J_c$ in single-band superconductors.  In the multi-band systems such as the 
iron-based superconductors their relations become much more complicated  and are still controversial \cite{kogan}. 
But, in any case the parameter $\gamma_\lambda$ describes the scale of the superconducting anisotropy. 
Thus, in this paper we perform a systematic first-principles study for $\gamma_\lambda$ in the whole 
groups of iron-based superconductors and elucidate the origin of the general trend in the superconducting 
anisotropy observed in various measurements such as torque, microwave surface impedance, $J_{\rm c}$, 
etc. \cite{prozorov,tanatar,kubota,weyeneth,weyeneth2,jaroszynski,kano,bukowski,jia,okazaki,martin}. 
Though accurate values of $\gamma_\lambda$ have not yet been settled experimentally, i.e., 
the reported values are still scattered, we find a clear trend in  $\gamma_\lambda$ among the five groups 
of the iron-based superconductors. 
Moreover, since we can predict the anisotropy ratio of a superconductor in which single crystals are not 
available,  our method presented in this paper will provide a powerful tool for the exploration of the 
superconducting materials suitable for each application.   
Our calculations reveal that  the anisotropy ratio $\gamma_\lambda$ depends sensitively on the nature 
of the blocking layers and its values are widely distributed as seen in experiments, e.g.,  
$\gamma_\lambda\sim 3$ in the superconductors belonging to the 111 and 122 groups and 
$\gamma_\lambda\gtrsim200$ in those with perovskite-based blocking layers. These results remind us of 
the variety in high-$T_{\rm c}$ cuprates, 
e.g., $\gamma_\lambda \sim 8$ in YBa$_2$Cu$_3$O$_{8-\delta}$ and $\gamma_\lambda > 100$ 
in Bi$_2$Sr$_2$CaCu$_2$O$_{8-\delta}$(Bi-2212) \cite{cuprate}. 
It is noted that the anisotropy in the 111 and 122 groups is small in spite of the layer structure.  The  
large value of $\gamma_\lambda$ in the group with perovskite-based blocking layers strongly suggests 
the formation of the intrinsic Josephson junction stacks along the $c$-axis.

In this paper, we evaluate $\gamma_\lambda$ for the typical iron-based superconductors from numerical 
results for the anisotropy ratio of the normal state resistivity at $T=0$K, $\gamma_\rho (0)$ obtained 
by means of the first-principles calculations for the electronic states \cite{hamada}, that is, the 
superconducting anisotropy originating from the band-structure dependent Fermi velocities is considered. 
\begin{table}[t]
\caption{Lattice constants $a$ and $c$, the height $h_{\rm Pn}$ of As, P , or Se ions from the Fe-plane, 
and the anisotropy parameters of the resistivity and the penetration depth at zero temperature.
$a$, $c$, and $h_{\rm Pn}$ are experimental values as the input parameters for the first-principles 
calculations, and $\gamma_\rho$ and $\gamma_\lambda$ are the calculated ones. 
}\label{tab}
\begin{ruledtabular}
\begin{tabular}{ccccccc}
         & $a$[\AA] & $c$[\AA] & $h_{\rm Pn}$[\AA] & $\gamma_\rho (0) $ & $\gamma_\lambda (0) $ \\
         \hline
FeSe     & 3.7738   & 5.5248   & 1.4652            & 18.44            & 4.29             \\
LiFeAs   & 3.7914   & 6.3639   & 1.6769            &  9.06            & 3.01             \\
BaFe$_2$As$_2$
         & 3.9625   &13.0168   & 1.3602            & 10.69            & 3.27             \\
LaFePO   & 3.9636   & 8.5122   & 1.1398            & 17.34            & 4.16             \\
LaFeAsO  & 4.020    & 8.7034   & 1.3238            &116.8             &10.81             \\
Sr$_2$ScFePO$_3$
         & 4.016    &15.543    & 1.1984            & $6.19\times10^5$ &248               \\
\end{tabular}
\end{ruledtabular}
\end{table}
Let us first summarize the method employed in this paper to evaluate the anisotropy parameters, 
$\gamma_\rho$ and $\gamma_\lambda$.   
Assuming that the relaxation time of the conduction electrons is 
isotropic and independent of their velocities, we consider the quantity 
$
\gamma_\rho(0) = {\langle v_a^2 \rangle_{\rm FS}}/{\langle v_c^2 \rangle_{\rm FS}}
$
to be equal to the ratio of the $c$-axis resistivity to the $a$-axis one. Here, $v_a$ and 
$v_c$ are the Fermi velocities, respectively, parallel and perpendicular to the FePn layers 
and $\langle\cdots\rangle_{\rm FS}$ denotes the average on the Fermi surface \cite{hamada}.
The Fermi velocities are derived from the derivative of the band energy, 
$
v_i = {\partial \epsilon_k}/{\partial k_i},
$
where the band energy $\epsilon_k$ is calculated by the first-principles calculations. 
In our calculations we use the first-principles density functional package VASP\cite{vasp}, which 
adopts GGA exchange-correlation energy \cite{pbe} and PAW method \cite{paw}.
The electron self-consistent loops to obtain the charge density are repeated until the total 
energy difference becomes smaller than $10^{-6}$ eV. In these loops the spacing between 
nearest-neighbor $k$-points is taken to be $\sim 0.1$--$0.2$ ${\rm \AA}^{-1}$. 
Once the charge density is determined, the energy bands are again calculated for more $k$-points 
located with a much finer spacing, 79$\times$79, in the $xy$-plane (the number of $k$-points along 
the $z$-axis depends on the compounds), in order to determine the Fermi surfaces and the Fermi 
velocities.
The average $\langle \cdots \rangle_{\rm FS}$ is calculated by the standard tetrahedron method.
In obtaining the anisotropy of the penetration depth at zero temperature, we utilize the simplest 
relation, $\gamma_\lambda (0) =\sqrt{\gamma_\rho(0)}$ \cite{kogan2}, which is convenient for 
systematic and comparative studies on the superconducting anisotropy among various iron-based 
superconductors, though its validity is limited in the multi-band systems\cite{kogan2}.
The compounds that we examine in this paper are FeSe, LaFeAsO, LaFePO, BaFe$_2$As$_2$, LiFeAs,
and Sr$_2$ScFePO$_3$. The calculations are done for tetragonal undoped compounds without any 
magnetic order, except for BaFe$_2$As$_2$. We calculate the electronic structures based on the 
observed crystalline structures, that is, the crystalline structures are not optimized, since 
it is known that the optimized structure deviates from the observed one in iron-based 
superconductors.  \cite{margadonna,tapp,rotter,lafepo,cruz,ogino}.

\begin{figure}
\includegraphics[scale=0.25]{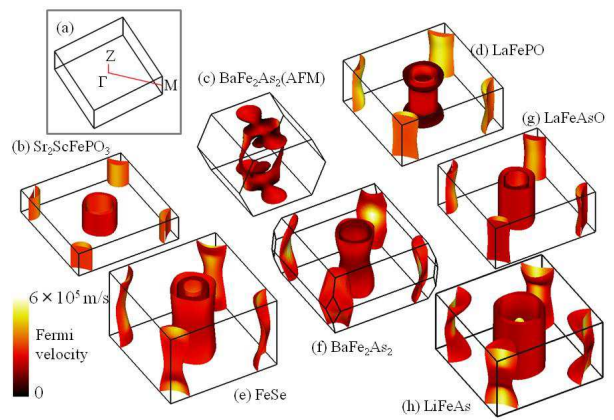}
\caption{The Fermi surfaces of (b) Sr$_2$ScFePO$_3$, (c) antiferromagnetic BaFe$_2$As$_2$, (d) LaFePO, (e) FeSe,
(f) BaFe$_2$As$_2$, (g) LaFeAsO, and (h) LiFeAs. The contrast on the surfaces indicates the magnitude of the Fermi velocity. 
The inset (a) shows typical symmetry points. }
\label{fig:fermi}
\end{figure}

The numerical results for the anisotropy parameters, $\gamma_\rho$ and $\gamma_\lambda$, 
in the iron-based superconductors employed in this paper are summarized in Table \ref{tab}. 
Let us discuss the 1111-system, first. The experimental values in this group are widely 
distributed, depending on the compounds
\cite{prozorov,kubota,weyeneth,weyeneth2,jaroszynski,jia,okazaki,martin}. 
For example, $\gamma_\lambda\sim 15-20$ in ReFeAsO (Re=Nd, Sm)
\cite{prozorov,martin,weyeneth2}, while  $\gamma_\lambda\sim 3$ in PrFeAsO \cite{okazaki}.
Thus, we notice that there are two groups in the 1111-system, i.e., the one with relatively large anisotropy and 
the other with moderate anisotropy. In our calculations, we obtained $\gamma_\lambda (0)=10.81$ for 
LaFeAsO, as shown in Table \ref{tab}. This value is close to that in the group with relatively large anisotropy. 
We also performed the calculations for stoichiometric PrFeAsO and obtained 
$\gamma_\lambda(0)\sim 8.6$, which is smaller than that in LaFeAsO as expected and also 
not far from the observed one in PrFeAsO$_{1-y}$.
This difference between LaFeAsO and PrFeAsO can be mainly attributed to the 
difference of the main elements inside the blocking layers.  
On the other hand, it is also noted that 
the calculated value in non-arsenic 1111-compound LaFePO ($\gamma_\lambda (0)=4.16$) is 
much smaller than that of LaFeAsO, though the crystal structures are equivalent in both compounds.  
The Fermi surface shapes are different in these two compounds as seen in Figure 
\ref{fig:fermi} (d) and (g), that is, the Fermi surfaces in LaFeAsO are almost cylindrical, 
while those in LaFePO are not, indicating that the two-dimensionality is weak in LaFePO. 
 From these results, one understands that the two-dimensionality differs among the 1111 
compounds even if the blocking layers are equivalent. This remarkable feature comes from 
the fact that the band dispersion along the $z$-direction is sensitively related to the Pn's height 
from the Fe's square lattice plane. 
It is also noticed that the anisotropy of the 1111 compounds ($\gamma_\lambda\gtrsim 10$) is 
on the same order as in YB$_2$Cu$_3$O$_{8-\delta}$, that is, the transport application will be 
promising in the 1111-compounds.

In the 122-compound BaFe$_2$As$_2$, we obtained $\gamma_{\lambda} (0) = 3.27$. This value is 
smaller than that of LaFeAsO (1111-system) and is consistent with the experimental one $\sim 6$ \cite{prozorov}, 
which is generally smaller than that in the 1111-system. 
The weak anisotropy in the 122-compound is also understood from the feature of the Fermi 
surfaces. As seen in Fig.\ref{fig:fermi}, the Fermi surfaces in this compound are more winding 
along the $z$-direction compared  with that in LaFeAsO, which leads to the weak anisotropy in the 
superconducting state of the 122-compounds.  This indicate that 
the 122 system has high potential for transport applications.
The 122-compound makes a tetragonal-to-orthorhombic transition, and a stripe-type 
antiferromagnetic (AFM) order appears in the low temperature phase. 
We also calculated the anisotropy parameter $\gamma_\rho$ in the antiferromagnetic phase ($\mu_{\rm Fe} =1.9 \mu_{\rm B}$)
\cite{note}
and found $\gamma_\rho(0) = 1.2$, which is smaller than that of the tetragonal one being 
consistent with the experiments. In fact, the experimental value is $\gamma_\rho \sim 2.5$ \cite{tanatar}.
Thus, one finds 
that the anisotropy becomes weaker in the AFM phase. Note that the reduction in the resistivity 
anisotropy originates from the three-dimensional Fermi surfaces appearing in the AFM phase as 
seen in Fig.2(c). These results clearly demonstrate the advantage of the present evaluation technique. 
In Table \ref{tab} we also list the anisotropy parameters of LiFeAs in the 111-system. The  
$\gamma_{\lambda}$ value is comparable to BaFe$_2$As$_2$, that is, the 111-system is also suitable 
for transport applications. This low anisotropy basically reflects the winding features of the 
Fermi surface like 122-systems as seen in Fig.2(h).

In the 11-compounds FeSe, which has the simplest crystal structure composed of a stack of only 
FePn layers, the Fermi surfaces calculated in this compound shows clear curvature along the 
$z$-direction as seen in Fig.2(e), indicating that the two-dimensionality is weak in this system. 
In fact, we obtained a relatively small value, $\gamma_\lambda (0) = 4.29$, in the superconducting 
state. However, we note that this value is larger than those in the 122 and 111 systems and rather close to that in LaFePO. 
 From this result one understands that metallic ions in the blocking layers enhance the three dimensionality 
in the 122 and 111 systems. We emphasize that only the first-principles calculations can systematically derive such 
delicate material-dependent difference in the anisotropy.  

\begin{figure}
\includegraphics[scale=0.18]{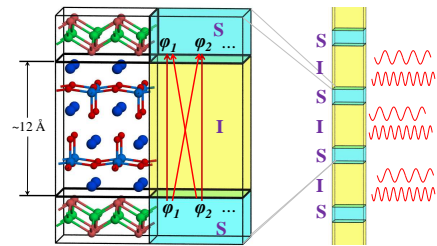}
\caption{A schematic illustration of the intrinsic Josephson junction stacks in Sr$_2$ScFePO$_3$.
An alternate stack of FePn (superconducting) and blocking layers (insulating) are regarded as a series 
of S-I-S  Josephson junctions.
$\varphi_i$ stands for the phase of the superconducting order-parameter of the $i$th electron energy band. 
The arrows represent possible tunneling channels.
}
\label{fig:joseph}
\end{figure}

Now,  we discuss the anisotropy in the final group, i.e.,  the newly discovered iron-based superconductors 
with perovskite-type blocking layers, which are now under intensive exploration. 
The superconductors in this group do not show 
any magnetic order at all,  while the superconducting transition temperature is rather high. 
In this paper we focus on a typical  compound, Sr$_2$ScFePO$_3$, which shows the highest $T_{\rm c}$ at 
ambient pressure among non-arsenic iron-based superconductors.
Since the blocking layers in this compound are very thick  as shown in Fig.1,  strong anisotropy  is expected in 
the normal state  resistivity. In fact, a huge value $\gamma_\rho\sim 10^{6}$ is obtained, which is  several  hundred 
times larger than that in the other groups,i.e.,  1111-, 122-, and 11- compounds.  From this normal state value 
it follows, $\gamma_{\lambda} (0) \sim 250 $,  for the anisotropy in the London penetration depth of Sr$_2$ScFePO$_3$. 
Note that the value is comparable to that in Bi-2212 in high-$T_{\rm c}$ cuprates. From this fact one may infer 
that the $c$-axis transport in this system is brought about by the electron tunneling between  neighboring FePn 
layers, as in Bi-2212. In the superconducting state the Cooper-pair tunneling will be also  expected to occur 
between the superconducting FePn layers as in Bi-2212,  that is,  a single crystal of Sr$_2$ScFePO$_3$ may be 
regarded as a stack of nano-scale Josephson junctions as schematically shown in Fig.\ref{fig:joseph}, which is 
called the intrinsic Josephson junctions. 
In the intrinsic Josephson junctions the $I-V$ characteristics under no external magnetic field show a remarkable 
feature, i.e., the {\it multiple branch structure} composed of many $I-V$ curves \cite{kleiner}, the number of which 
corresponds to that of stacked intrinsic SIS junctions as shown in Fig.\ref{fig:joseph}.  Hence, we naturally 
expect that the multiple-branch structure appears in the $c$-axis $I-V$ characteristics  in the superconducting 
state in a single crystal of Sr$_2$ScFePO$_3$. 
As for the damping nature in the Josephson effects, this system should be under-damped one, because the paring symmetry 
is expected to be mainly full gapped $s$-wave one. 
It is also predicted that a new oscillation mode exists in the phase differences in addition to the Josephson plasma 
\cite{otaprl}. Moreover, the Josephson vortex has an internal structure \cite{ota2}.
These new features originates from the multi-tunneling channels
due to the multi-gap nature in these superconductors as shown in Fig.\ref{fig:joseph}.
The existence of the new phase oscillation modes will enrich the physics in the 
intrinsic Josephson junctions expected in the new compounds. 
We also mention that the superconducting Sr$_2$ScFePO$_3$ is a stoichiometric compound, that is, the 
superconductivity appears without doping. Then, it will be easier to make a stack of  homogeneous intrinsic 
Josephson junctions in a large-scale, which is contrasted with the Bi-2212 intrinsic Josephson junctions, 
in which the doping for getting the high-$T_{\rm c}$ superconductivity brings about disorders.

\begin{figure}
\includegraphics[scale=0.45]{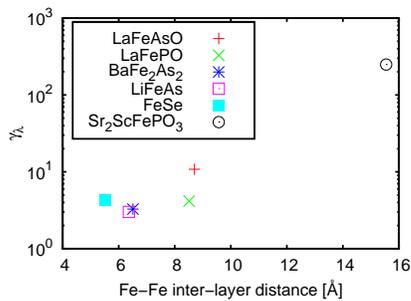}
\caption{The penetration-depth anisotropy as a function of the distance between  Fe-planes.}
\label{fig:c-gamma}
\end{figure}

Finally, we summarize our numerical results for $\gamma_\lambda$ vs. distance between adjacent iron-planes 
in  Fig.\ref{fig:c-gamma}.  From this figure one finds a rough trend between the distance and the anisotropy,  
which indicates that the anisotropy increases with increasing the distance.  However, our results also suggests the 
existence of the other factors that affect the penetration depth anisotropy. 
As discussed in this paper, the anisotropy is sensitive to the distance between the pnictogen ions and the Fe-plane, 
and also the elements forming the blocking layers. From these results 
we claim that the method presented in this paper, using the first-principles calculations, 
is a powerful tool for evaluating the superconducting anisotropy.

In summary, we investigated systematically the anisotropy of the London penetration depth for various iron-based 
superconductors on the basis of the first-principles calculations for their electronic states.  It was shown that the 
anisotropy parameters, $\gamma_\rho$ and $\gamma_\lambda$, are strongly dependent on the structure of the 
blocking layers and pnictogen ions situated next to Fe layers. Our numerical results well explain the variety of 
the $\gamma_\lambda$ values observed in these superconductors. 
The 122- and the 111-compounds show weak anisotropy, being enough for transport power applications, 
while those with perovskite-type blocking layers have a very large value of $\gamma_\lambda$, which is comparable 
to that in Bi-2212. We predict that the intrinsic Josephson junction stacks are realized in the superconducting 
state of these compounds.

The authors wish to thank H. Ogino and J. Shimoyama 
for providing their recent experimental results and K. Terakura 
 for illuminating discussion in first-principles calculations. 
 The authors also thank N. Hayashi, Y. Nagai, M. Okumura, and N. Nakai 
 for valuable discussion. The work was partially supported by Grant-in-Aid 
 for Scientific Research on Priority Area ``Physics of new quantum phases 
 in superclean materials" (Grant No. 20029019) from the Ministry of 
 Education, Culture, Sports, Science and Technology of Japan.


\begin{thebibliography}{99}
\bibitem{note}
The calculated moment becomes larger than the observed
one. This is a well-known problem in first-principles calculations for
iron-based superconductors.  We avoid further discussion on this
discrepancy. See e.g., \cite{mazin} for its more details.
\end{thebibliography}
\end{document}